# MIDOG 2025: Mitotic Figure Detection with Attention-Guided False Positive Correction

Andrew Broad, Jason Keighley, Lucy Godson, Alex Wright: *National Pathology Imaging Cooperative*, UK

## Abstract

We present a novel approach which extends the existing Fully Convolutional One-Stage Object Detector (FCOS) for mitotic figure detection. Our composite model adds a Feedback Attention Ladder CNN (FAL-CNN) model for classification of normal versus abnormal mitotic figures, feeding into a fusion network that is trained to generate adjustments to bounding boxes predicted by FCOS. Our network aims to reduce the false positive rate of the FCOS object detector, to improve the accuracy of object detection and enhance the generalisability of the network. Our model achieved an F1 score of 0.655 for mitosis detection on the preliminary evaluation dataset.

MIDOG 2025 | Artificial Intelligence | Neural Networks | Attention | Mitotic Figure Detection

Contact: andrew.broad@nhs.net

## Introduction

Accurate mitosis detection in histopathological images is an important task in cancer grading and prognosis. However, the variability in staining, tissue morphology, and mitotic figure (MF) appearance across datasets present challenges in automating this task using Artificial Intelligence (AI) methods. The MIDOG 2025 challenge therefore aims to find generalizable solutions for mitotic segmentation and grading (Ammeling et al., 2025). The work in this submission addresses Track 1 of the challenge, mitosis detection.

We present a novel pipeline combining existing models, the Feedback Attention Ladder Convolutional Neural Network (FAL-CNN) (Broad et al., 2024) and the Fully Convolutional One-Stage Object Detector (FCOS) (Tian et al., 2019). We adopt a unique fusion approach to enhance the positioning and confidence score of the bounding boxes (bboxes) predicted by the FCOS. We use the FAL-CNN classifier to generate a spatial attention map and a probability of mitosis ($p_{mitosis}$) for mini-patches centred on the FCOS-predicted bounding boxes. These results are processed in a downstream fusion network which is trained to generate modified bounding box locations and scores, in the hope of fine-tuning these results for greater accuracy in MF detection and rejecting false-positive MFs. The composite model is designed as a drop-in replacement for the baseline FCOS object detector.

## Material and Methods

**Our image processing pipeline** integrates four main components:

- **FCOS object detector** (Tian et al., 2019): A fully convolutional one-stage anchor-free object detector that predicts bounding boxes and class scores from an input image of the region of interest (ROI).
- **FAL-CNN classifier** (Broad et al., 2024): A hierarchical feedback attention-based CNN, trained to classify mitotic and non-mitotic figures and to generate spatial attention maps aligned with cellular features of interest, to guide feature extraction.
- **Fusion network**: Combines spatial attention map and mitotic figure probability to generate coordinate offsets and score multipliers, used to fine-tune the position and score of the FCOS-predicted bounding boxes

- **Mini-patch sampling:** The bounding box locations predicted by the FCOS model are used as the centres of 56x56px samples taken from the input image. These are batched and used as the input to the FAL-CNN classification stage.

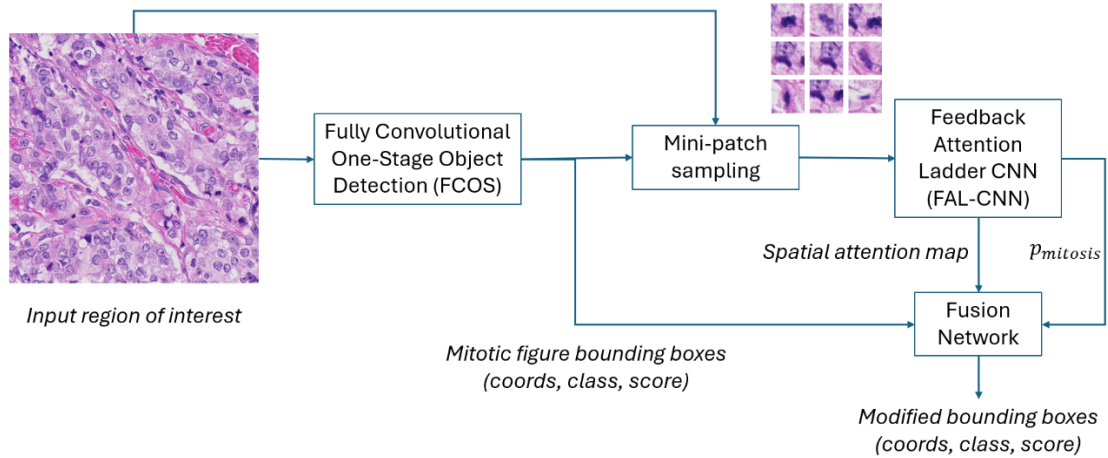

Figure 1: Mitotic Figure Detection with Attention-Guided False Positive Correction

Baseline methods provided by the MIDOG 2025 organisers include the use of the FCOS object detection model. The network was trained using an NVIDIA A40 GPU.

The FAL-CNN consists of a feedforward CNN based on VGG19 (Simonyan and Zisserman, 2015), with a symmetrical feedback pathway which generates spatial attention maps which weight the feedforward path to enhance the model's response to salient image regions. The attention maps are also supplied as a model output which can be used in object location.

The fusion network concatenates the coordinates, class and score, from one FCOS-predicted bounding box at a time, with the mitosis probability and the high-level (14x14) spatial attention map from the FAL-CNN resulting from processing image content defined by the original bounding box. This combined input is passed through three fully connected linear layers with output sizes 48, 12 and 3, alternating with rectified linear unit (ReLU) layers. The fusion network outputs bbox position offsets and score-multiplier values which are used to adjust the bbox from the FCOS output.

**Data Sources:** The MIDOG 2025 challenge only allows for the use of open source or publicly available datasets. Within this work we used the MIDOG++ dataset (Aubreville et al., 2023) for training of the Feedback attention ladder and fusion network. For training the FAL-CNN, 56x56px patches were extracted from MIDOG++ training images (Aubreville et al., 2023), with patch centres aligned with ground truth bbox annotations. A balanced split was used, with 11,937 mitotic figure and 11,937 non-mitotic figure patches. Datasets were split into three groups, a training group (70%), a testing group (20%) and a validation group (10%).

**Training:** The FCOS model was trained using 553 region of interest images and 26,287 annotated mitotic figure locations in the MIDOG++ challenge data (Aubreville et al., 2023), using gradient descent and maximizing F1 score agreement over 150 epochs with learning rate (LR) reduction x0.7 per 30 epochs with initial LR=1e-4. The FAL-CNN was trained to classify MF and non-MF patches, with a similar LR schedule starting from LR=1e-5, to minimize cross entropy loss over 50 epochs. The Fusion network was trained *in situ* in the composite model, for 150 epochs and using pre-trained FCOS and FAL-CNN networks with frozen weights, otherwise using the training configuration previously used for FCOS alone.

Results

The model yielded an F1 score of 0.655 when uploaded to MIDOG 2025 for the Track 1 Evaluation phase. The baseline FCOS model, prior to adding the FAL-CNN or fusion model and trained on the same MIDOG++ data (Aubreville et al., 2023), achieved F1=0.767.

## Discussion

The solution proposed within this work aims to add an extension to existing object detection networks, allowing for more complex networks with the FAL-CNN and fusion network checking and reclassifying based on secondary knowledge.

Within the initial test phase, this algorithm failed to outperform the baseline FCOS algorithm. However, our expectation is that the model will be less affected by shifts in datasets and domains. Final analysis on unseen test sets will prove or disprove this hypothesis. We were only able to evaluate one prototype fusion model for our submitted architecture. Further development here might reveal a more effective way of fine-tuning the FCOS model's prediction to enhance F1 performance. The FAL-CNN generates feedback attention maps at multiple resolutions, which can be used to control local saccade-like movements to re-sample patches to centre them on cellular features relating to mitotic figures. We believe that further work is justified to explore the diagnostic benefits of this behaviour.

Our FAL-CNN feedback mechanism has further application in explainability: The feedback attention maps can highlight salient features associated with MF/non-MF classifications, with the potential to detect diseased cells. We propose further work to explore this as a diagnostic enhancement to the valuable bounding box information already returned by the FCOS model.

## Bibliography

1. Ammeling, J., Aubreville, M., Banerjee, S., Bertram, C. A., Breininger, K., Hirling, D., Horvath, P., Stathonikos, N., & Veta, M. (2025). Mitosis Domain Generalization Challenge 2025. Medical Image Computing and Computer Assisted Intervention 2025 (MICCAI). Zenodo. https://doi.org/10.5281/zenodo.15077361.
2. Aubreville, M., Bertram, C.A., Donovan, T.A., Marzahl, C., Maier, A., Klopfleisch, R., 2020. A completely annotated whole slide image dataset of canine breast cancer to aid human breast cancer research. Scientific Data 7, 417. https://doi.org/10.1038/s41597-020-00756-z
3. Aubreville, M., Wilm, F., Stathonikos, N., Breininger, K., Donovan, T.A., Jabari, S., Veta, M., Ganz, J., Ammeling, J., Van Diest, P.J., Klopfleisch, R., Bertram, C.A., 2023. A comprehensive multi-domain dataset for mitotic figure detection. Scientific Data 10, 484. https://doi.org/10.1038/s41597-023-02327-4
4. Broad, A., Wright, A., McGenity, C., Treanor, D., de Kamps, M., 2024. Object-based feedback attention in convolutional neural networks improves tumour detection in digital pathology. Sci Rep 14, 30400. https://doi.org/10.1038/s41598-024-80717-3
5. Simonyan, K., Zisserman, A., 2015. Very Deep Convolutional Networks for Large-Scale Image Recognition. https://doi.org/10.48550/arXiv.1409.1556
6. Tian, Z., Shen, C., Chen, H., He, T., 2019. FCOS: Fully Convolutional One-Stage Object Detection, in: 2019 IEEE/CVF International Conference on Computer Vision (ICCV). Presented at the 2019 IEEE/CVF International Conference on Computer Vision (ICCV), pp. 9626–9635.